\magnification=1200
{\nopagenumbers
\baselineskip=15pt
\hsize=5.5in
\hskip10cm UM--P-96/74\hfil\par
\hskip10cm RCHEP-96/07 \hfil\par
\hskip10cm Revised Version \hfil 
\vskip2cm
\centerline{\bf Higher Twist Effect in Inclusive Quarkonium Photoproduction}
\vskip2cm
\hskip4cm J.P. Ma  \par

\hskip4cm Recearch Center for High Energy Physics \par
\hskip4cm School of Physics\par
\hskip4cm University of Melbourne \par
\hskip4cm Parkville, Victoria 3052\par
\hskip4cm Australia \par
\hskip4cm E-Mail: ma@physics.unimelb.edu.au \par
\vskip2cm
{\bf\underbar{Abstract}}:\par
We analyze higher twist effect in photoproduction
of quarkonium, where the quarkonium is a spin-triplet, S-wave
state. We find that the nonperturbative effect of next-to-leading twist
is contained in three correlation functions related
to the initial hadron and some of the effect of 
next-to-leading twist is only suppressed by the inverse of the mass square of
the quarkonium. A naive estimation indicates that the effect can be 
significant for charmonium. 
In the analysis we only take the leading order in the 
small velocity expansion for the nonperturbative parts 
related to the quarkonium. Possible corrections from higher 
orders, especially, corrections from color-octet states 
are discussed in detail. 
\par\vskip15pt
\noindent
PACS Numbers: 12.38.Bx, 13.85.Ni, 14.40.Gx. 
\par
\noindent
Keywords: Twist-4 effect, quarkonium production.  
\par\vfil\eject}
\hsize=5.5in
\baselineskip=15pt
\pageno=1
\noindent 
{\bf 1. Introduction}
\vskip 15pt
Recently the theory of quarkonium production has been developed 
rapidly(See recent reviews given in [1]). 
Quarkonium production can be regarded as a two-step
process, where a heavy quark $Q$ and its antiquark $\bar Q$ 
are produced and then this $Q\bar Q$ pair is transmitted 
into a quarkonium. The production of the $Q\bar Q$ pair 
can be treated with perturbative QCD because the mass $M_Q$ 
of $Q$ is large, while the transition is a nonperturbative 
process. It is realized that in a rest frame of the quarkonium 
the quark $Q$ or the antiquark $\bar Q$ moves with a small 
velocity $v$. Hence a small velocity expansion can be used 
to treat the nonperturbative transition. As a result the effect
of the transition can be described by quarkonium matrix elements 
of local operators defined in nonrelativistic QCD[2]. 
It is interesting to note that not only a color-singlet $Q\bar Q$, 
but also a color-octet $Q\bar Q$ can be transmitted into a quarkonium. 
This is in contrast to the color-singlet model and is specially of 
importance to P-wave quarkonium production, because the production rates
through a color-singlet and through a color-octet are at the same order
of $v$. 
For $S$-wave quarkonium, although the contributions from 
a color-octet $Q\bar Q$ is suppressed at least by $v^2$ in comparison
with those from a color-singlet, they can be 
significant[1]. Besides this nonperturbative effect related to the transition 
there is another nonperturbative effect if there are hadrons 
in the initial state. This nonperturbative effect can be analyzed 
with a twist expansion. At the order of leading twist it results 
in the well-known parton model. The subject of the present work 
is to analyze the effect from the order of next-to-leading twist 
in inclusive photoproduction of quarkonium.    
\par\vskip 5pt
In the case of inclusive photoproduction of a single jet with a    
large transversal momentum $P_T$, one can expect that higher 
twist effects are suppressed by the power of $P^{-1}_T$, since $P_T$ 
is the only large scale in the problem. If one observes instead of
a single jet a quarkonium state, there is, in addition to $P_T$, 
another large scale--the mass of the heavy quark $M_Q$. 
Naturely one expects that some effects from higher twist 
may be only suppressed by the power  
of $M_Q^{-1}$. In this work we assume that no polarization is
observed, so the powers will be even numbers.     
We find that some effect of next-to-leading
twist is suppressed indeed by $M_Q^{-2}$. This indicates that 
the effect from higher twist may be substantial 
in the production of charmonium, as the mass $M_c$ is not
so large. 
\par\vskip5pt
In this work we only consider the quarkonium 
as a S-wave, spin-triplet state and denote it as $\psi$. For charmonium
and bottonium it corresponds to $J/\psi$ and to $\Upsilon$ respectively. 
We will assume that the nonperturbative parts related to the
quarkonium and to the initial hadron can be factorized separately 
and analyze in detail the effect of next-to-leading 
twist in the $\psi$-production, where the formation 
of $\psi$ from a $Q\bar Q$ pair is considered 
at leading order of $v^2$. At this order, only a color-singlet
$Q\bar Q$ can be transmitted into $\psi$. 
At higher orders of $v$ there are relativistic corrections and 
contributions from 
color-octet states of $Q\bar Q$ pair.
We will discuss
this in length in a separate section, where we also discuss
electroproduction briefly. 
To analyze the nonperturbative effect related to 
the initial hadron we employ 
the diagram expansion method. This method was first used to classify
higher twist effects in deeply inelastic scattering[3], it  
exactly delivers the same results as those obtained 
with traditional Wilson operator expansion[4]. 
This method was also applied or extended to other processes.
An incompleted list of references can be found in [5-10]. 
However, the extension of the method to our case is not quite 
straightforward. In deeply inelastic scattering, because the
process is totally inclusive, the perturbative parts of 
diagrams with different numbers of incoming glue as partons 
can be related to each other with the help
of low energy theorems[3], basically derived from the Ward identity. 
This leads to that the contributions from diagrams with a given
number $N$ of incoming glue and from diagrams with $N+1$ incoming 
glue can be combined together by replacing a space-time
derivative in the corresponding nonperturbative 
part with a gauge covariant derivative. Hence $SU(3)$ gauge symmetry
is maintained explicitly. The process we consider is seminclusive, 
where a $Q\bar Q$ pair is observed indirectly, thus the corresponding  
low-energy theorems are lacking. Further, the perturbative
parts are much more complicated than those in deeply
inelastic scattering. In this work we will only consider 
in the diagram expansion those diagrams whose perturbative 
parts are at the same order of coupling constants and will simply 
replace space-time derivatives in nonperturbative parts by
gauge covariant derivatives.  
\par\vskip5pt
In our work we will assume that we do not observe any 
polarization of any particle, and we take only leading order in  
coupling constants. We will also assume that the produced $\psi$ has 
moderate transverse momentum, i.e., the 
$\psi$ is not in the forward region in a fixed target 
experiment. In this region various processes can 
produce $\psi$, some of them can not be described 
with perturbative QCD. We neglect heavy flavor content in the initial hadron, 
which is expected to give small contribution. 
An interesting aspect of the processes is that
at this order only glue are responsible for the quarkonium production.
Studying these processes will help to understand the role played by gluon 
in hadrons. The process considered here was used to study 
gluon distribution in nucleon before(See [11] and references cited there).    
\par\vskip5pt
Our work is organized as the following: In Sect.2 we introduce
notations and the diagrams in the diagram expansion which
are needed to be considered. The results at the leading twist is 
also given. In Sect.3 we work out the results
for next-to-leading twist. A discussion of our results is presented. 
In Sect.4 we will discuss possible contributions 
from higher orders of $v$, where  
production of a color-octet $Q\bar Q$ pair can also lead to contributions 
for the total production rate. In this section electroproduction 
is also discussed briefly. 
We summarize 
our work in Sect.5.  
\par\vskip20pt
\noindent
{\bf 2. The Diagram Expansion and Results of Leading Twist}
\par\vskip15pt
We consider the inclusive process:
 $$ \gamma(q)+H(P) \rightarrow \psi(k) +X. \eqno(2.1)$$
The momenta of particles are given in the brackets. We choose a frame
in which the hadron $H$ moves in the $z$-direction and the photon in
the opposite direction. The photon is on-shell and $H$ has mass $M_H$. 
Throughout our work we will neglect $M_H$. The correction introduced
by the target mass may be included by using Nachtmann variable[12]. 
In our approximation the quarkonium mass $M_\psi$ is twice of $M_Q$. 
We define $s=(q+P)^2$ and $t=(P-k)^2$. 
For performing the analysis it is convenient to use light-cone coordinate
system. In this system a 4-vector is expressed as $p^\mu =(p^+, p^-,p^1,p^2)
=(p^+,p^-, {\bf p_T})$. The relation to conventional expression is 
$p^+={1\over \sqrt{2}}(p^0+p^3),\ p^-={1\over \sqrt{2}}(p^0-p^3)$. In the 
system the photon carries the momentum $q^\mu =(0, q^-,0,0)$ and $H$ 
$P^\mu =(P^+, {M_H^2\over 2P^+}, 0,0)$. We introduce in the frame 
two vectors $n$ and $\ell$ and a tensor $d_T^{\mu\nu}$: 
  $$ \eqalign{ n^\mu & =(0,1,0,0),\ \ \ \ell^\mu =(1,0,0,0),\cr
    d_T^{\mu\nu} &= g^{\mu\nu}-n^\mu\ell^\nu -n^\nu\ell^\mu.\cr
   } \eqno(2.2) $$ 
Contracting any vector with $d_T^{\mu\nu}$ gives its transverse part. 
We will work in the light-cone gauge $n\cdot G(x)=G^+(x)=0$, where $G^\mu(x)
=G^{a, \mu}(x)T^a$
is the gluon field. In this gauge a component of the gluon 
field strength tensor $G^{\mu\nu}$ takes a simple form: 
  $$ G^{a,+\mu}(x)= {\partial \over \partial x^-} G^{a,\mu} (x). \eqno(2.3) $$  
\par\vskip10pt
To analyze the effect upto next-to-leading twist we need to   
consider the diagrams given in Fig.1A, Fig.1B and Fig.1C in the diagram expansion
approach. In these diagrams  the black boxes represents the nonperturbative
part involved by the hadron $H$, the upper parts, i.e., 
the blank boxes,  contain the perturbative 
part for production of the $Q\bar Q$ pair at the leading order of 
coupling constants and the transition of the pair into $\psi$, 
they represent the sample of Feynman diagrams given in Fig.2A, Fig.2B and Fig.2C
correspondingly.   
These diagrams are with Cutkosky cut since we are interested in cross
section. With twist-classification, Fig.1A has the leading twist 2, 
Fig.1B has 3 and Fig.1C has 4. The contribution from Fig.1A can be written as:
  $$ \eqalign { 2s d \sigma_A &= {d^4k\over (2\pi )^4} 2\pi\delta
    (k^2-M^2_\psi ) \int {d^4p_1\over (2\pi )^4} 
     A^{ab}_{\mu\nu}(p_1)  \cr 
  &\ \ \ \ \ \ \ \cdot \int d^4 x e^{-ix\cdot p_1} 
   <H(P)\vert G^{a,\nu}(x) G^{b,\mu}(0)\vert H(P)> \cr } \eqno(2.4)$$
In Eq.(2.4) $A^{ab}_{\mu\nu}$
corresponds to the upper part in Fig.1A, this part is the contribution 
from diagrams given in Fig.2A. 
The hadronic matrix element in Eq.(2.4) 
corresponds to the black box in Fig.1A and it is averaged over the spin
of $H$. Because of the color symmetry we can write Eq.(2.4) as
  $$\eqalign { 2s d \sigma_A &= {d^4k\over (2\pi )^4} 2\pi\delta
    (k^2-M^2_\psi )      
   {1\over 8} \int {d^4p_1\over (2\pi )^4}
     S_{\mu\nu}(p_1) \cr 
   &\ \ \ \ \ \ \ \cdot \int d^4 x e^{-ix\cdot p_1}
   <H(P)\vert G^{b,\nu}(x) G^{b,\mu}(0)\vert H(P)> \cr
     S_{\mu\nu}(p_1)  &=  A^{aa}_{\mu\nu}(p_1) \cr }  \eqno(2.5)$$
\par 
The contribution from Fig.1C takes the form:
 $$ \eqalign { 2s d \sigma_C &= {d^4k\over (2\pi )^2} 2\pi\delta
   (k^2-M^2_\psi ) \int {d^4k_1 \over (2\pi)^4}{d^4k_2\over (2\pi)^4} 
     {d^4k_3\over (2\pi)^4} C^{a_1a_2a_3a_4}_{\mu_1\mu_2\mu_3\mu_4} 
    (k_1,k_2,k_3) \cr 
    & \cdot \int d^4x_1  d^4 x_2 d^4 x_3 e^{-ix_1\cdot k_1
    -ix_2\cdot x_2 +ix_3\cdot k_3 } \cr  
    & \ \ \ \ \ <H(P)\vert G^{a_1,\mu_1}(x_1) G^{a_2,\mu_2} (x_2) G^{a_3,\mu_3}(x_3)
       G^{a_4,\mu_4}(0) \vert H(P)>  \cr} \eqno(2.6) $$ 
where $C$ is the upper part and corresponds to the diagrams given in Fig.2C. 
Note $C$ and $A$ are at the same order of $\alpha$ and $\alpha_s$.  
At the leading twist the momenta $k_i'$s in the perturbative part $C$ 
will be set to be proportional to $P$, so the quarkonium will be produced
with zero transversal momentum. We will not consider this type of contribution
as mentioned in the introduction.  
As for the contribution from Fig.1B, the perturbative part 
is one order   
higher in $g_s$ than those in Fig.1A. Again we will not consider it. 
It is worth to point out that some contributions from diagrams at higher 
order of $g_s$ will be included in our final results, because
we later replace space-time derivatives with gauge covariant derivatives
for  gauge invariance. 
In the case we consider the contribution to the process 
is only from Fig.1A with the perturbative part specified 
with the Feynman diagrams given in Fig.2A. 
The main task for analyzing twist-4 effect is to separate
them from the contribution from Fig.1A. 
\par\vskip10pt
To achieve the separation one observes that the space-time dependence
of the matrix elements in Eq.(2.5) is controlled by 
different scales 
in different directions. In $-$-direction this scale is $P^+$ and it 
is very large. In $+$-direction and transverse direction 
this scale is just $P^-$ or $\Lambda_{QCD}$, the $\Lambda$-
parameter of QCD,  
which are small. With this observation one can approximate
the $x^+$- and ${\bf x_T}$-dependence by Taylor-expanding $x^+$ and ${\bf x_T}$. 
This expansion is equivalent to the expansion of $S_{\mu\nu}(p_1)$
around $p_1^{\mu}=\hat p_1^{\mu} =(p^+_1,0,0,0)$. The expansion reads:
  $$ \eqalign {S_{\mu\nu} (p_1) &= S_{\mu\nu} (\hat p_1) 
   +{\partial S_{\mu\nu} (p_1) \over \partial p_1^\alpha }
   \vert_{p_1=\hat p_1} \Delta p_1^\alpha \cr 
  &\ \ \  +{1\over 2} {\partial^2 S_{\mu\nu} (p_1) \over \partial
    p_1^\alpha \partial p_1^\beta } \vert_{p_1=\hat p_1} \Delta p_1^\alpha
    \Delta p_1^\beta +\cdots \cr } \eqno(2.7) $$
where $\Delta p_1 =p_1-\hat p_1$. The $\cdots$ denotes higher order in 
$\Delta p_1$ which will lead contributions with twist higher than 4.  
The expansion is called as collinear expansion. The leading twist 
contribution comes only from the first term in the expansion.  
This term does not depend on $p_1^-$ and ${\bf p_{1T}}$.    
Taking this term for $S_{\mu\nu}$ in Eq.(2.5) we can perform 
the integration over $p_1^-$ and ${\bf p_{1T}}$ directly: 
   $$ \eqalign { 2s d \sigma_A &={d^4k\over (2\pi )^4 }2\pi\delta
   (k^2-M^2_\psi )
     {1\over 8}\int {dz\over z} S_{\mu\nu}(\hat p_1) \cr
    & \cdot {zP^+ \over 2\pi }\int dx^- e^{-izP^+x^-} 
     <H(P)\vert G^{b,\nu}(x^-n) G^{b,\mu}(0) \vert H(P)> \cr
    & +"{\rm Higher\ Twist \ Terms} " \cr} \eqno(2.8) $$
where we made the substitution $p_1^+ =zP^+$. Because of the symmetries
of Lorentz boost along $z$-direction and of rotation around $z$-axis, 
the Fourier transformed matrix element can be written: 
   $$ \eqalign { {zP^+\over \pi}  \int dx^- e^{-izP^+x-} 
    & <H(P)\vert  G^{b,\nu}(x^-n) G^{b,\mu}(0) \vert H(P)>= \cr &  d_T^{\mu\nu} 
     {zP^+ \over 2\pi } \int dx^- e^{-izP^+x^-}
   <H(P)\vert G^{b,\rho}(x^-n) G^{b} _{\ \rho} (0) \vert H(P)> \cr
    & +n^{\mu}n^\nu {zP^+ \over \pi}   
   \int dx^- e^{-izP^+x^-} 
     <H(P)\vert G^{b,-}(x^-n) G^{b,-}(0) \vert H(P)>, \cr }\eqno(2.9)$$ 
where $d_T^{\mu\nu}$ is given in Eq.(2.2).
In light-cone gauge the dynamically independent gluon fields
are $G^1$ and $G^2$, $G^-$ can be solved with the equation of motion 
and the term in Eq.(2.9) with $G^-$ will be shown in the next section 
to be at higher twist. The leading twist contribution from Fig.1A
can be written now as: 
  $$ \eqalign { 2s d \sigma_A &= {d^4k\over (2\pi )^4} 2\pi\delta
   (k^2-M^2_\psi ) 
    \int {dz\over z} f_{G/H}(z) 
    \cdot [- {1\over 16 }d_T^{\mu\nu} \cdot S_{\mu\nu} (\hat p_1) ] 
    \cr 
   & + "{\rm Higher\ Twist \ Terms }" \cr} \eqno(2.10) $$ 
The term in $[\cdots ]$ is just the probability for the photon-gluon fusion 
into $\psi$ and a gluon, where the initial gluon is on shell because of
$\hat p_1^2 =0$. Therefore the interpretation of parton model applies. 
The function $f_{G/H}(z)$ can be read from Eq.(2.9): 
   $$ f_{G/H}(z)=-{zP^+ \over 2\pi }\int dx^- e^{-izP^+x^-}
     <H(P)\vert G^{b,\rho}(x^-n) G^{b} _{\ \rho} (0) \vert H(P)>. 
     \eqno(2.11)$$
Because of the total momentum conservation in $+$ direction the variable 
$z$ is constrained between 0 and 1. 
The function  $ f_{G/H}(z)$ is positive and can be 
interpreted as the probability of finding a gluon 
in $H$ with the energy fraction $z$. It is preferred to use strength tensor
instead gluon field to define $f_{G/H}(z)$. Using Eq.(2.3) one can obtain:
  $$ f_{G/H}(z)= {-1 \over 2\pi zP^+} \int dx^- e^{-izP^+x^-}
   <H(P)\vert G^{b,+\rho }(x^-n) G^{b,+} _{\ \ \ \rho} (0) \vert H(P)> 
   \eqno(2.12) $$ 
This is the definition of gluon distribution given in [13] in light-cone 
gauge. For arbitrary gauge one needs to supply Wilson line operator
between the two tensor operators. 
\par\vskip5pt
The perturbative result at the leading twist is well known. For completeness
we will give this result. For this purpose we introduce
   $$ \eqalign{  \hat s &=(q+\hat p_1)^2,\ \ \ \ \hat t=(k-\hat p_1)^2, \cr 
      \sigma &= \sigma ^{(0)} +\sigma^{(2)} +\cdots . \cr} \eqno(2.13)$$
$\sigma ^{(0)}$ denotes the contribution at leading twist. $\sigma^{(2)}$
stands for the contribution at twist-4. The invariant $\hat s$ and $\hat t$ 
by neglecting $M_H$ are simply related to $s$ and $t$ via $\hat s=zs$ 
and $(\hat t -M^2_\psi )  =z(t-M^2_\psi)$ respectively. The result at leading twist is:
  $$ \eqalign { {d\sigma^{(0)} \over dt} &= {32
       \pi \over 3} \alpha \alpha_s^2
    Q^2_Q M_\psi \vert R_\psi (0)\vert ^2 \int {dz\over z}  f_{G/H}(z)\cr 
    & \ \ \cdot  { \hat s^2 (\hat s-M_\psi^2)^2 +\hat t^2 (\hat t-M_\psi^2)^2 
     +(\hat s+\hat t)^2(\hat s +\hat t -M_\psi^2)^2 
    \over \hat s^2 (\hat s-M_\psi^2)^2 (\hat t -M_\psi^2)^2  
       (\hat s+\hat t)^2 } , \cr}\eqno(2.14)$$ 
where $Q_Q$ is the electric charge of $Q$ in unit $e$. $R_\psi (0)$ is the radial 
wavefunction for $\psi$ at the origine, which is related to a matrix element
defined in [2]. 
\par\vskip5pt
The contribution from the next-to-leading twist is from the term with 
$n^\mu n^\nu$ in Eq.(2.9) and the terms with derivatives in Eq.(2.7).   
We will analyze them in the next section. 
\par\vskip 10pt
\vfil\eject

\noindent 
{\bf 3. The Results at the Next-to-leading Twist}
\par\vskip 15pt
We start first with the term with $n^\mu n^\nu$ in Eq.(2.9). This term corresponds
to the contribution from the nonpropagating part of the gluon propagator[5]. 
In this term only the $-$-components of gluon fields are involved and they are 
not dynamically independent. With the equation of motion in QCD for these
component one can solve them in terms of transverse components of gluon fields
and quark fields. The equation of motion reads:
  $$ {\partial G^{a,+-}(x)\over  \partial x^-}= -(D_T^{\mu} G^+_{\ \mu}(x))^a -g_s J^{a,+}
     (x), \eqno(3.1) $$
where $D_T^{\mu}=d_T^{\mu\nu}D_\nu$ and $D_\nu$ is the covariant derivative in   
the adjoin representation. 
$J^{a,+}$ is the $+$-component of quark color current. 
Assuming all fields approach zero at infinite space-time, one can solve 
this equation formally:
  $$  G^{a,+-}(x^+,x^-,{\bf x_T}) =-{1\over 2} \int^{+\infty }_{-\infty}
     d\xi \tau (x^- -\xi ) \{(D_T^{\mu} G^+_{\ \mu}(x^+,\xi,{\bf x_T}))^a 
      +g_s J^{a,+}(x^+,\xi,{\bf x_T})\}. \eqno(3.2) $$           
The function $\tau (z)$ equals 1 for $z>0$ and -1 for $z\le 0$. To use this 
solution we write $G^-$ into $G^{+-}$  in the second term in Eq.(2.9) 
with help of Eq.(2.3) by partial 
integration.
Finally we obtain the contribution from this
term: 
  $$  2sd\sigma_{A1}  = {d^4 k\over (2\pi)^4} (2\pi) 
      \delta (k^2-M_\psi^2)  
     \int {dz \over z} e_1(z) \{ {1\over 8} 
      { n^\mu n^\nu \over (n\cdot \hat p_1)^2} S_{\mu\nu}(\hat p_1) \}  
  \eqno(3.3) $$
where the function $e_1(z)$ is defined as
  $$ \eqalign { e_1(z)  ={1\over 2\pi zP^+} \int dx^- e^{-izP^+ x^-}& 
    <H(P)\vert [ (D_T^{\mu} G^+_{\ \mu}(x^-n))^a +g_s J^{a,+}(x^-n)] \cr 
  & \ \ \   
    \cdot [ (D_T^{\nu} G^+_{\ \nu}(0))^a +g_s J^{a,+}(0)]    
   \vert H(P)> .     \cr } \eqno(3.4) $$
This function is positive and has  support only for $z$ between 0 and 1.
It has a dimension 2 in mass while $ f_{G/H}(z)$ is dimensionless. 
Therefore this term will lead to contribution at order of next-to-leading twist.  
We used subscript $A1$ in Eq.(3.3) to denote this contribution. The   
perturbative part in this contribution is the part in $\{\cdots \} $ in 
Eq.(3.3) and can be calculated straightforward. We obtain:
 $$ \eqalign { {d\sigma _{A1} \over dt} &= 
    {32\pi \over 3} \alpha \alpha_s^2
    Q^2_Q M_\psi \vert R_\psi (0)\vert ^2 \int {dz\over z} 
     \cdot {e_1(z)\over M^2_{\psi} } 
     \cdot {1\over \hat s^3 (\hat s-M_\psi^2)^2 (\hat t -M_\psi^2)^2
       (\hat s+\hat t)^2 } \cr 
   & \cdot \big\{ 
  2\hat s^3(\hat s +\hat t)^2 -4M_\psi^2 (\hat s^4 +\hat s^3 \hat t 
      +2\hat s \hat t^3 +2\hat t^4 ) 
     +2M_\psi^4 (\hat s^3 +\hat s\hat t^2 +4\hat t^3) \big\} \cr } 
   \eqno(3.5) $$ 
\par
Comparing this contribution with that from the leading twist in Eq.(2.14)
there is  an extra $M_\psi^{-2}$ term, the contribution hence may be 
suppressed by this power. 
\par\vskip10pt
Next we consider the contribution from the terms with derivatives 
in the collinear expansion in Eq.(2.7). The second term in Eq.(2.7) 
leads a contribution which can be written as:
  $$ \eqalign { {d^4 k\over  (2\pi)^4} & (2\pi)
      \delta (k^2-M_\psi^2) 
            \int {d^4p_1\over (2\pi )^4}
     {1\over 16} \big [ { \partial S_{\mu\nu} \over 
      \partial p_1^\alpha } \vert _{p_1=\hat p_1}\big ] \cr
   & \cdot \ \ (-i) \int d^4 x e^{-ix\cdot p_1} 
   < H(P) \vert (\partial^\alpha G^{b,v}(x)) G^{b,\mu} (0) 
    +G^{b,\nu}(0) (\partial^\alpha G^{b,\mu}(-x)) \vert H(P)> ,
   \cr } \eqno(3.6) $$
where we have used partial integration and translational invariance
for the hadronic matrix element. Since we do not observe any spin
of any particle, by using P(Parity)- and 
T(Time reversal)- symmetries one can show that: 
  $$ S_{\mu\nu}(p_1) = S^{\nu\mu} (p_1), \eqno(3.7)$$ 
while the matrix element is antisymmetric in $\mu\nu$ in contrast to 
Eq.(3.7). Hence  
the second term in Eq.(2.7) does not contribute. 
If any polarization is observed, this term will lead to a 
twist-3 effect.  
\par\vskip5pt
The third term in Eq.(2.7) leads to a contribution given by: 
$$ \eqalign {   
     \int {d^4 k \over (2\pi)^4} & (2\pi) \delta (k^2-M^2_\psi) 
         \int {d p_1^+ \over 2\pi } {1\over 16 } 
   \big [ {\partial ^2 S_{\mu\nu} (p_1) \over 
    \partial p_1^\alpha \partial p_1^\beta } \vert _{p_1=\hat p_1} 
    \big ] \cr
  & \ \cdot (-1) \int dx^- e^{-ix^- p_1^+} 
   < H(P) \vert G^{b,\nu} (x) (\partial^\alpha \partial^\beta 
        G^{\mu} (0) )^b \vert H(P) > . \cr } \eqno(3.8) $$
It should be noted that by using the translational invariance
one can let the derivatives acting on the gluon fields differently
than that given above and hence there is an ambiguity 
to form operators in the matrix element. We take the above form 
since the operator is just the trace term subtracted in 
 the corresponding
twist-2 operator. The leading contribution 
in Eq.(3.7) comes from those terms in which the contracting    
Lorentz indices $\mu,\ \nu, \alpha$ and $\beta$ are only 
the transverse indices. If any of them is $-$, one can show 
that the contribution is at higher twist. For $\mu$ or $\nu$  being 
$-$ the argument is the same as for the second term in Eq.(2.8). 
For $\alpha$ or $\beta$ in the matrix element 
being $-$ one can show because of the covariance
of Lorentz boost along $z$-direction that the corresponding
contribution is suppressed by the factor ${1 \over P^+}$     
relative to the terms with $\alpha$ or $\beta$ being transverse 
indices. 
Similarly as to Eq.(2.9) one can write the Fourier 
transformed matrix element in Eq.(3.8) as: 
  $$ \eqalign { {-p_1^+ \over 2\pi } & \int dx^- e^{-ix^- p_1^+} 
   < H(P) \vert G^{b,\nu} (x) (\partial^\alpha \partial^\beta 
        G^{b,\mu} (0) ) \vert H(P) > \cr 
      & = (d_T^{\mu\alpha} d_T^{\nu\beta} +d_T^{\mu\beta}      
           d_T^{\nu\alpha} -d_T^{\mu\nu} d_T^{\alpha\beta} ) 
   \cdot d_1(z)  \cr
    & \  + (3d_T^{\mu\nu}d_T^{\alpha\beta} 
     -(d_T^{\mu\alpha} d_T^{\nu\beta} +d_T^{\mu\beta}      
           d_T^{\nu\alpha}) )\cdot d_2(z) \cr
     & \ \ \ +"{\rm High\  Twist\  Terms }" \cr} \eqno(3.9) 
  $$ 
where the functions $d_1(z)$ and $d_2(z)$ are defined as
  $$ \eqalign { d_1(z) &= {-1 \over 2\pi zP^+} 
     \int dx^- e^{-izP^+ x^-} 
     {1\over 4} <H(P) \vert 
    G^{a,+}_{\ \ \ \sigma} (x^-n) (D_T^{ \{ \sigma } D_T^{\rho \} }          
     G^{+}_{ \ \rho} (0) )^a \vert H(P)> ,\cr
      d_2(z) & = {-1 \over 2\pi zP^+} 
     \int dx^- e^{-izP^+ x^-} 
     {1\over 8} <H(P) \vert 
    G^{a,+}_{\ \ \ \sigma} (x^-n) (D_T^2 
     G^{+}_{ \ \sigma } (0) )^a \vert H(P)>. \cr} \eqno(3.10) $$
In Eq.(3.9) we replaced derivatives with covariant derivatives 
and $\partial_T^\sigma \partial_T^\rho$ with 
$D_T^{ \{ \sigma } D_T^{\rho \} }$. The symbol $\{ \sigma \rho \}$
means to take the part symmetric in $\sigma\rho$. 
Calculating moments of the distributions $d_1(z)$ and $d_2(z)$ one can
show that the moments correspond the matrix elements of local operators,
which are the trace-terms subtracted in the twist-2 gluonic operators. 
We denote the twis-4 contribution from Eq.(3.8) to the 
cross-section as $\sigma_{A2}$, which is:
 $$ \eqalign { 2sd\sigma_{A2} & = {d^4k \over (2\pi)^4} (2\pi)
        \delta (k^2-M_\psi^2)  
  \int {dz\over z}  \cr  
  & \  \cdot      \Big \{  
    d_1(z) \cdot \big [ {1\over 16} \cdot 
     {\partial ^2 S_{\mu\nu} (p_1) \over
    \partial p_1^\alpha \partial p_1^\beta } \vert _{p_1=\hat p_1}
      \big ] \cdot (d_T^{\mu\alpha} d_T^{\nu\beta} +d_T^{\mu\beta}
           d_T^{\nu\alpha} -d_T^{\mu\nu} d_T^{\alpha\beta} ) \cr
      & \  + d_2(z)\cdot \big [ 
     {1\over 16} \cdot 
     {\partial ^2 S_{\mu\nu} (p_1) \over
    \partial p_1^\alpha \partial p_1^\beta } \vert _{p_1=\hat p_1}
      \big ] 
     \cdot (3d_T^{\mu\nu}d_T^{\alpha\beta}
     -(d_T^{\mu\alpha} d_T^{\nu\beta} +d_T^{\mu\beta}
           d_T^{\nu\alpha}) ) \Big \} .\cr} \eqno(3.11)  $$    
\par
Some care must be taken for calculating this part. In $S_{\mu\nu}(p_1)$
there is a $\delta$-function $\delta ((q+p_1-k)^2)$, which is the 
on-shell condition for the gluon crossing the cut in Fig.2A. 
Therefore the derivatives of $S_{\mu\nu}(p_1)$ 
are constrained derivatives. To calculate the derivatives we first perform
the integration of $z$ with the $\delta$-function in Eq.(3.11), 
where ${\bf p_{1T}}$ is not zero. After the integration $p_1^+$ 
is fixed 
and it depends 
on ${\bf p_{1T}}$. Now the derivatives can be easily
performed, where the indices $\alpha$ and $\beta$ are transversal 
indices. Finally we rewrite the result as an integration of $z$ 
with the $\delta$-function and set $p_1=\hat p_1$.  
The calculation is tedious but straightforward.
The results for the perturbative part
at order of $\alpha\alpha_s^2$ are:
$$ \eqalign { {d\sigma _{A2} \over dt} &=
    {32\pi \over 3} \alpha \alpha_s^2
    Q^2_Q M_\psi \vert R_\psi (0)\vert ^2 \int {dz\over z}
   \cdot {1\over \hat s^2 (\hat s-M_\psi^2)^4 (\hat t -M_\psi^2)^4
      (\hat s+\hat t)^2} \cr
   &  \ \ \ \ \ \ \ \ \  \cdot \big \{
    f_1(\hat s,\hat t) \cdot {d_1(z) \over M_\psi ^2}
       +f_2( \hat s,\hat t)
      \cdot {d_2(z) \over \hat s}
     \big \}, \cr} \eqno(3.12) $$
where the functions $f_1$ and $f_2$ are:
$$ \eqalign { f_1(\hat s, \hat t) =& 8\hat s^4 \hat t^2 (\hat s+\hat t)^2 
    +16M^2_\psi \hat s\hat t (-\hat s^5 -5\hat s^4\hat t
     -4\hat s^3\hat t^2 +2\hat s^2\hat t^3 +6\hat s\hat t^4 +4\hat t^5) \cr
    & +8M^4_\psi (\hat s^6 +10\hat s^5 \hat t +21\hat s^4 \hat t^2
      -8\hat s^3\hat t^3 -48\hat s^2\hat t^4 -40\hat s\hat t^5 
      -8\hat t^6 )+16M^6_\psi (-\hat s^5 
      \cr & -9\hat s^4 \hat t
      +\hat s^3 \hat t^2  
       +35 \hat s^2\hat t^3 +40\hat s\hat t^4 
     +14\hat t^5)+ 8M^8_\psi (-2\hat s^4+8\hat s^3\hat t 
   -37\hat s^2\hat t^2-78\hat s\hat t^3 
      \cr & -43\hat t^4) 
      +16M^{10}_\psi (3\hat s^3 +14 \hat s\hat t^2 +17\hat t^3) 
      +8M^{12}_\psi (-3\hat s^2+2\hat s\hat t -11\hat t^2) , \cr
      f_2(\hat s, \hat t) =& +32\hat s^2\hat t(-2\hat s^5-\hat s^4\hat t
     +2\hat s^3\hat t^2 +2\hat s^2\hat t^3 -2\hat s\hat t^4
     -3\hat t^5) +16M^2_\psi\hat s(-2\hat s^6 +6\hat s^5\hat t
      \cr
    & -3\hat s^4\hat t^2-6\hat s^3\hat t^3+5\hat s^2\hat t^4 
        +16\hat s\hat t^5 +2\hat t^6 )+32M^4_\psi ( 
      4\hat s^6 +\hat s^5\hat t+2\hat s^4\hat t^2 -\hat s^3\hat t^3
      \cr & -8\hat s^2\hat t^4
       -3\hat s\hat t^5-\hat t^6) 
       +32M^6_\psi(-6\hat s^5 -2\hat s^4\hat t +\hat s^3\hat t^2
     +3\hat s^2\hat t^3 +2\hat s\hat t^4+3\hat t^5) 
     \cr & +32M^8_\psi (4\hat s^4-\hat s^3 \hat t -\hat s^2\hat t^2
     +\hat s\hat t^3-3\hat t^4) +32M^{10}_\psi (-\hat s^3+\hat s^2\hat t
   -\hat s\hat t^2 +\hat t^3). \cr} \eqno(3.13) $$  
\par\vskip5pt
At next-to-leading twist the contribution to the total
cross-section for the process in Eq.(2.1) is the sum:
  $$ \sigma^{(2)}=\sigma_{A1} +\sigma_{A2}. \eqno(3.14) $$ 
In our results the nonperturbative parts at order of twist-4 in the 
process are the correlation functions $e_1(z)$, $d_1(z)$ 
and $d_2(z)$, while that at order of leading twist is $f_{G/H}(z)$. 
These functions are universal, i.e., they will appear in other 
processes at certain order of $\alpha_s$, provided 
the same hadron $H$ is in the initial state. The functions 
$e_1(z)$, $d_1(z)$ and $d_2(z)$ have a dimension  
2 in mass and they are at order of $\Lambda^2_{QCD}$. The 
perturbative parts corresponding to these functions
are calculated at leading order of $\alpha$ and $\alpha_s$ in this 
section. Comparing the results with that in the last section 
we can see that the effect from next-to-leading twist 
is suppressed by the inverse of one scale which can be
$\hat s$, $\hat t$ or $M^2_\psi$, or any linear combination of them. 
To see more clearly how the twist-4 effect is suppressed we consider 
two limiting cases, these cases can be realized with suitable cuts in experiment. 
One case is 
the kinematical region where $\hat s,\ \vert \hat t\vert >> M_\psi^2$, 
but $\hat t$ is not so large that the $\psi$ does not appear in the forward
region, whose definition will be given later. 
Another is for $\hat s >> \vert \hat t\vert, M_\psi^2$.  
\par\vskip10pt
Case 1). $\hat s,\ \vert \hat t\vert >> M_\psi^2$. In this case, 
one can neglect any power of ${M_\psi^2 \over \hat s}$ and ${M_\psi^2 
\over  \hat t}$. In this approximation the expression 
for the differential cross-section simplifies and the contribution related
to $d_2(z)$ can be neglected:
 $$ \eqalign { {d\sigma \over dt} &= {32
       \pi \over 3} \alpha \alpha_s^2
    Q^2_Q M_\psi \vert R_\psi (0)\vert ^2 \int {dz\over z}  \cdot
     {1\over \hat s^2 \hat t^2 } \cr 
 & \  \cdot \big \{ 
   {\hat s^4 +\hat t^4 +(\hat s+\hat t)^4 \over \hat s^2 (\hat s+\hat t)^2} 
    \cdot f_{G/H}(z) +2{e_1(z)\over M_\psi^2}   
   +8 {d_1(z)\over M_\psi^2 }
  \big\} .\cr} \eqno(3.15) $$  
From Eq.(3.15) we clearly see that the twist-4 effect is suppressed 
only by $M^{-2}_\psi$, but not by the inverse of 
any large scale of $\hat s$ or $\hat t$. 
\par\vskip5pt
Case 2). $\hat s >> M_\psi^2, \vert\hat t\vert$. In this limit we neglect 
any power of ${M_\psi^2 \over \hat s}$ and ${\hat t \over\hat s}$. 
The differential cross-section becomes
  $$ \eqalign { {d\sigma \over dt} &= {32
       \pi \over 3} \alpha \alpha_s^2
    Q^2_Q M_\psi \vert R_\psi (0)\vert ^2 \int {dz\over z}  \cdot
     {1\over \hat s^2 (\hat t-M^2_\psi)^2 } \cr 
   & \ \cdot \big\{ 2 f_{G/H}(z) +2{e_1(z) \over M_\psi^2} 
     +8{d_1(z)\over M_\psi^2 }-32 {(M^2_\psi +2\hat t) \over (\hat t 
    -M^2_\psi )^2 }\cdot d_2(z)   \big\} .\cr }\eqno(3.16) $$
Again the twist-4 effect here is suppressed by $M_\psi^{-2}$ or by 
the inverse of a linear conbination of $M^2_\psi$ and $\hat t$. 
\par\vskip5pt
In general there is always some higher twist effect which is suppressed 
only by $M^{-2}_\psi$. From the discussions of the above two cases 
one can not make this effect negligible through 
suitable cuts in experiment. Therefore the twist-4 effect may be substantial
for charmonium production, as the mass $M_c$ is not so large. 
Because the nonperturbative functions $e_1(z)$, $d_1(z)$
and $d_2(z)$ are unknown, one does not know how  large the twist-4 effect
can be. A naive estimation could be given by noting
the fact that the effect we considered is basically from the correction
of the parton-model, where one takes the off-shellness and 
the transversal momentum of the parton into account. With 
this fact one may make an Ansatz for a naive estimation 
as follows:
   $$ e_1(z)=d_1(z)=d_2(z)=\Lambda^2 f_{G/H}(z), \eqno(3.17)$$
where the parameter $\Lambda$ can be regarded as an average 
of the transversal momentum of the initial gluon and it should be
at the order corresponding the radius of the initial hadron.   
With this Ansatz the differential cross-section upto 
twist-4 can be written:
    $$  {d\sigma \over dt } =  {32
       \pi \over 3} \alpha \alpha_s^2
    Q^2_Q M_\psi \vert R_\psi (0)\vert ^2 \int {dz\over z} 
     (w_2(\hat s, \hat t) +\Lambda^2 w_4(\hat s, \hat t))\cdot 
       f_{G/H}(z). \eqno(3.18) $$
We will give some numerical results for the quantity        
$\rho$ defined  as
  $$\rho ={w_2(\hat s, \hat t) +\Lambda^2 w_4(\hat s, \hat t)
       \over w_2(\hat s, \hat t)}. \eqno(3.19)$$ 
If the twist-4 effect is negligible, $\rho$ should be close to 1.
An useful variable used commonly in experiment is $z$, it is defined  
as:  
    $$ z={k\cdot P\over q\cdot P}. \eqno(3.20)$$
which should not be confused with the intergration variable $z$ 
in previous equations. If $\psi$ is produced with $z>0.9$,
it is in the froward region. 
We plot the quantity $\rho$ as a function of $z$ for a
given $\hat s$, 
where we take $\Lambda =500$MeV, this value
may correspond to a proton as the initial hadron. 
The results are given in Fig.3 for $\hat s =(10{\rm GeV})^2$
and for $\hat s =(30{\rm GeV})^2$ respectively and the quarkonium
is $J/\psi$.  
On can take these as representative estimations 
for $\sqrt s=30$GeV and $\sqrt s =100$GeV.  
From the figure we note that the twist-4 effect 
becomes smaller when $\hat s$ becomes larger and it is more significant
in lower $z$-region than that in higher $z$-region. 
For $\hat s=(30{\rm GeV})^2$ it can be an effect at $30\%$. 
\par\vskip20pt
\noindent 
{\bf 4. Higher Order Effect in the Small Velocity Expansion}
\par\vskip15pt
In previous sections we analyzed the nonperturbative effect
at twist-4 related to the initial hadron, while the 
nonperturbative effect related to the quarkonium is only considered
at leading order of the small velocity expansion. In this
section we discuss higher order effect in this expansion. 
At leading order of $v$ only a color-singlet $Q\bar Q$ pair 
in a $^3S_1$ state can be transmitted into the $^3S_1$
quarkonium, the probability for this can be regarded as 
$\vert R_\psi (0)\vert ^2$, which should be related 
to a matrix element defined in NRQCD[2]. The relation is
  $$ \vert R_\psi (0)\vert ^2 = {2\pi \over 9} <0\vert 
      O^\psi _1 (^3S_1) \vert 0> . \eqno(4.1)$$ 
We will not give the definition of the matrix element
and of those appearing later. For the definitions we refer 
reader to [2]. We will take the leading order here as 
$v^0$. Corrections to the leading order results will 
start at order of $v^2$.  
\par\vskip 5pt 
At next-to-leading order, i.e., at $v^2$, there is only 
the relativistic correction, which takes the effect of the 
relative motion of $Q$ and $\bar Q$ into account. 
This effect is analyzed 
within the color-singlet model in [14], where the correction 
is parameterized in term of the parameter $\epsilon$ defined  
as:
  $$ M_\psi =2M_Q+\epsilon.   \eqno(4.2)$$
Numerical results given in [14] show that for charmonium the correction 
at $\sqrt s =14.7$GeV is significant in the high $z$ region, 
where $z$ is defined in Eq.(3.20) and $\epsilon=0.16M_c$. 
However, there is a uncertainty  
in choosing the parameter $\epsilon$.  If one thinks that 
the binding force is Couloumb force, $\epsilon$ defined in Eq.(4.2) 
is negative. 
One can also show by starting the energy-momentum tensor the
relation  
that $\epsilon$ equals to the negative matrix element in 
NRQCD of the kinetic-energy  operator of $Q$ or
$\bar Q$ inside the quarkonium, if $M_Q$ in Eq.(4.2) is
the pole-mass[15]. However, this relation holds at the tree-level
and at the leading order of $v$. All these show that 
$\epsilon$ can also be negative. With a negative $\epsilon$ 
the correction turns into negative too. It should  be pointed
out that the analyze in [14] is  done with color-singlet model.
With the new NQRCD factorization in [2] the nonrelativistic
correction is parameterized with other type of
matrix elements instead of $\epsilon$ and a re-analyze 
seems needed. 
\par\vskip 5pt
At order of $v^4$, except the relativistic corrections there are also
other corrections due to the fact that   
a quarkonium state
consists many components, where a $Q\bar Q$ pair carrying the
quantum numbers of the quarkonium is the dominant one.
Other components consists fo a $Q\bar Q$ pair and soft glue, 
the probability to find them in a quarkonium is suppressed
by powers of $v$ relatively to that of the dominant one.  
Because of this a produced $Q\bar Q$ with quantum numbers
other than those of the quarkonium can become
one of these components by combining soft glue and
hence the production of such $Q\bar Q$ pair will 
also contribute to the production rate
of the quarkonium. In our case such $Q\bar Q$ pair can be a color-octet
$Q\bar Q$ pair in $^1S_0$, $^3S_1$ or $^3P_J$ state, where $J=0,1,2$.  
We use the notation $Q\bar Q_{(8)}(^{2S+1}L_J)$ to denote 
these states.   
With the velocity power counting rule given in [16] and 
the method in [2] one can show that the contributions from 
these color-octet $Q\bar Q$ pairs are at order of $v^4$. 
Taking the state $Q\bar Q_{(8)}(^3P_1)$ 
as an example, the probability to find such state combined 
with a soft gluon as a component of the quarkonium is at order
of $v^2$, while the formation of $Q$ and $\bar Q$ into a $P$-wave
state has a chance at order of $v^2$, hence 
the production of the quarkonium through such state 
will happen at order of $v^4$.  
\par
We first discuss the contribution from a $Q\bar Q_{(8)}(^3S_1)$ state. 
Such state can be produced not only by photon-gluon fusion
at order $\alpha\alpha^2_s$, 
but also by 
photon-quark scattering at the same order. 
However, numerical results show that the contribution from 
the photon-quark scattering is negligible.  
It is interesting to note that the contribution 
from the photon-gluon fusion is represented by the same
diagrams for the color-singlet contribution worked out before. 
To include this contribution one needs 
only to replace the matrix element in our
previous results via: 
  $$ <0\vert O^\psi _1 (^3S_1) \vert 0> 
 \rightarrow <0\vert O^\psi _1 (^3S_1) \vert 0>
   + {5\over 6}\cdot <0\vert O_8^\psi (^3S_1) \vert 0> . \eqno(4.3)$$ 
where the NRQCD matrix element $<0\vert O_8^\psi (^3S_1) \vert 0>$ 
is for the transition of the color-octet $Q\bar Q$. 
Due to the uncertainties in matrix elements and others like 
effects in higher order of $\alpha_s$, $K$-factor problem $\cdots$ etc, 
it is hard to identify this color-octet contribution. 
\par\vskip 5pt 
The contributions from the $Q\bar Q_{(8)}(^1S_0)$, $Q\bar Q_{(8)}(^3P_J)$ 
state at leading twist are analyzed in [17,18,19] for $J/\psi$. 
The matrix element for the transition of these states
into $J/\Psi$, i.e., $<0\vert Q_8^\psi(^1S_0)\vert 0>$ 
and $<0\vert O_8^\psi (^3P_J)\vert 0>$, is estimated[19,20].  The situation here 
is controversial. In the forward region, the contributions are 
from the subprocess $\gamma +G\rightarrow Q+\bar Q$ at the order 
of $\alpha \alpha_s$. 
In experiment the measured cross section 
contains also contributions  from elastic diffractive processes 
which can not be described with perturbative QCD. That means 
that the cross section     
estimated from the subprocess should be smaller than the measured one, 
if one neglects all higher order effects. A comparison with experiment
for $J/\psi$ production 
shows that the predicted cross section from the subprocess[17] 
and also including color-singlet contributions[18] is much larger 
than measured one, if one uses the numerical values for matrix 
elements estimated from experiment at Tevatron[20], they are 
roughly: 
  $$\eqalign {    <0\vert Q_8^\psi(^1S_0)\vert 0> & \sim 10^{-2} {\rm GeV}^3, \cr  
       {<0\vert O_8^\psi (^3P_0)\vert 0>
       \over M_c^2 } &= {1\over 2J+1}\cdot {<0\vert O_8^\psi (^3P_J)\vert 0>
       \over M_c^2 }  \sim 10^{-2} {\rm GeV}^3 . \cr } 
    \eqno(4.4)$$ 
It seems that these matrix elements are overestimated at least by 
one order of magnitude. Various reasons for that 
discrepancy are proposed like corrections from higher order of 
$\alpha_s$ and uncertainties in extracting 
numerical values of the matrix elements. One should also note
that the predictions may suffer large uncertainties because 
of twist-4 effect, as this effect is only suppressed 
by $M_Q^{-2}$ in the forward region. In addition, the relativistic 
correction at order of $v^2$ discussed before is significant 
in the froward region and can be negative.  
\par\vskip5pt
In the nonforward region, these color-octet states can be produced
at order of $\alpha\alpha_s^2$. At leading twist the production
is through photon-gluon fusion described by Fig.1A, where 
the blank part represents not only the Feynman diagrams given in Fig.2A, 
but also those like the one given in Fig.4, 
in which the gluon in the final state
is emitted by the incoming gluon. A comparison 
with available experiments is made in this region in [17,18]. It is 
shown that the contribution from these color-octet states 
becomes dominant for $z>0.5$ and including this contribution
leads to that the differential cross section increases rapidly
as $z$ increases. 
This rapid increasing however is 
not favoured by experimental data[17,18]. 
If the relativistic correction  discussed before 
is negative, the increasing may become mild. 
An important question related to our work is whether
there is some twist-4 effect suppressed only by $M_Q^{-2}$ 
as in the color-singlet contribution or not. 
The answer is no. To see this one needs to understand 
why the twist-4 effect analyzed before 
is suppressed by $M^{-2}_Q$. If one expands the 
perturbative part at leading twist in $M_Q$, the leading power
is 2. In the final result in Eq.(2.14) an extra $M_Q^{-1}$ comes in
because of the normalization. The leading power of the perturbative
parts at twist-4 is 0. Therefore there is twist-4 effect suppressed
only by $M_Q^{-2}$ in the production of a $^3S_1$ quarkonium. 
For the above color-octet states the leading power 
of the perturbative parts at leading twist is zero or negative,
and in calculating perturbative parts at twist-4 extra negative power
in $M_Q$ can not be generated, so there will be no twist-4 effect which 
is suppressed only by $M_Q^{-2}$ like those in Eq.(3.15).   
\par\vskip5pt
Before ending this section we briefly discuss electroproduction. 
This is basically the process $\gamma^* +H \rightarrow \psi +X$
where the photon $\gamma^*$ is off-shell and the off-shellness
is $-Q^2$. One can perform an analysis of twist-4 effect as we 
did here for a real photon. One can show for a fixed $Q^2$ 
that there is no twits-4 effect suppressed by $M_Q^{-2}$. 
The reason is the same as for the color-octet states mentioned
above. In photoproduction the forward region is the kinematical 
region where the produced quarkonium 
moves roughly in the same direction as the photon beam in a
fixed target experiment. In this region the contribution 
from elastic and diffractive processes to the production rate
is significant. In electroproduction like HERA experiment 
the produced quarkonium in the forward region   
can have large angle in respect to the hadron beam, one may 
expect that the contribution from elastic and diffractive processes
may not be significant. If the events in this region 
can be described 
by perturbative QCD, the theoretical predictions indicate
that there is a peak in the forward region, resulted by color-octet
states produced by the subprocess $\gamma^* +G \rightarrow 
Q+\bar Q$ and the subprocess at twist-4 described by Fig.1C. 
In experiment NMC data of $J/\psi$-production shows that there 
really is a peak
in the forward region[21]. However, the data was given as
the rapidity distribution, hence many information was lost for
a detail study of the peak. On the other hand, it is also
pointed out in the second reference in [11] that the peak can be well
explained by the theoretical results at leading twist.
If more data in experiment is available, this region 
may be a good place to study  the effect  from those
color-octet states and from twist-4. 
\par\vskip 20pt
\noindent 
{\bf 5. Summary} 
\par\vskip5pt 
In this work we have analyzed twist-4 effect in photoproduction
of a $^3S_1$ quarkonium. The nonperturbative parts at order of twist-4 are 
factorized into 3 correlation functions related to the initial hadron. 
The perturbative parts are calculated at leading order of coupling constants. 
We find that the twist-4 effect suppressed by $M^{-2}_Q $
in comparison with that at leading twist.
Hence the twist-4 effect
in photoproduction of $J/\psi$ may be substantial, and it can not
be made negligible by experimental cuts. 
Naive estimations indicate that the effect is significant at $\sqrt s$ 
around several ten GeVs. However, a definite conclusion requires 
a detailed study. 
\par
In the analysis we only take the leading order of the small velocity 
for the nonperturbative effect related to the quarkonium. The effects 
from higher orders of $v$ upto $v^4$ are discussed in detail. 
Especially, at order of $v^4$ production of various color-octet states 
of $Q\bar Q$ can contribute to the total production rate. However, 
the contributions from these states will not receive from twist-4 
corrections suppressed only by $M_Q^{-2}$ in the nonforward region, 
except the contribution from the color-octet state $Q\bar Q_{(8)}(^3S_1)$. 
But it is hard to distinguish this contribution from those
of the color-singlet as discussed in the last section.  
We have also discussed some aspects in electroproduction, here the 
twist-4 corrections are also not suppressed by $M^{-2}_Q $. 
It seems that the appearance of such suppression in photoproduction
of $^3S_1$ quarkonium may be a special case. 
\par
Unlike the gluon distribution $f_{G/H}(z)$, which is measured in experiment 
in many places, the three correlation functions for the nonperturbative
effect of twist-4 are unknown and need to be measured. 
These three functions are defined in Sect.3 in the light-cone gauge. 
In arbitrary gauge
one needs to supply Wilson-line operators between the gluonic operators 
in these definition. It should be noted 
that the definitions in Sect.3 are unrenormalized versions, i.e., at
tree level. A definition in renormalized version like that for 
correlation functions at twist-2 given in [13] may be difficult. The main
obstacles are that these functions may have power divergences and the 
perturbative parts corresponding to correlation functions
at twist-2 may have problem with convergence due to possible
presence of renormalon. The later indicates that there may be still 
some nonperturbative effect contained in the perturbative 
part at leading twist. A perfect definition would be given so that 
the two problems would be solved(See discussions in [22]). 
\par
Finally, it should be noted that the twist-4 effect in the kinematic 
region considered here is basically from the corrections of the 
parton model, where one takes the transverse momentum and 
the off-shellness of incoming partons into account. 
With this in mind it is possible to make a model 
for a more realistic estimation of twist-4 effect than the naive estimations
given in Sect.3. Currently a model is under development 
in hope to study the effect in detail. 
\par\vfil\eject  
\centerline{\bf References}
\par\vskip10pt
\noindent
[1] E. Braaten, S. Fleming and T.C. Yuan, OHSTPY-HEP-T-96-001, hep-ph/96-02374     
\par\noindent
\ \ \ \ M. Beneke, Talk given in the Second Workshop on Continous Advances in QCD, 
\par\noindent
\ \ \ \ hep-ph/9605462
\par\noindent
[2] G.T. Bodwin, E. Braaten and G.P. Lepage, Phys. Rev. D51 (1995) 1125
\par\noindent
[3] R.K. Ellis, W. Furmanski and P. Petronzio, Nucl. Phys. B212 (1983) 29
\par\noindent 
[4] R.L. Jaffe and M. Soldate, Phys. Lett. B105 (1981) 467, Phys. Rev. D26 (1982) 49
\par\noindent 
\ \ \ \ H.D. Politzer, Nucl. Phys. B172 (1980) 349  
\par\noindent
[5] J. Qiu, Phys. Rev. D42 (1990) 30
\par\noindent
[6] R.L. Jaffe and X. Ji, Nucl. Phys. B375 (1992) 527
\par\noindent
\ \ \ \ X.Ji, Nucl. Phys. B402 (1993) 217, Phys. Rev. D49 (1994) 114
\par\noindent
[7] J. Qiu and G. Sterman, Nucl. Phys. B353 (1991) 105, ibid 137 
\par\noindent
[8] M. Luo, J. Qiu and G. Sterman, Phys. Lett. B279 (1992) 377, 
\par\noindent 
\ \ \ \  Phys. Rev. D49 (1994) 4493 , {\it ibid} D50 (1994) 1951 
\par\noindent
[9] J. Levelt and P.J. Mulders, Phys. Rev. D49 (1994) 96
\par\noindent
[10] J.P. Ma, Nucl. Phys. B460 (1996) 109 
\par\noindent
\ \ \ \ \  H. Khan and P. Hoodbhoy, Phys. Rev. D53 (1996) 2534 
\par\noindent
[11] A.D. Martin, C.-K. Ng and W.J. Stirling, Phys. Lett. B191 (1987) 200
\par\noindent 
\ \ \ \ \ \ H. Merabet, J.-F. Mathiot and R. Mendez-Galain, Z. Phys. C62 (1994)
639
\par\noindent
[12] O. Nachtmann, Nucl. Phys. B63 (1973) 237 
\par\noindent
[13] J.C. Collins and D.E. Soper, Nucl. Phys. B194 (1982) 445
\par\noindent
[14] H. Jung et al. Z. Phys. C60, (1993) 721
\par\noindent
[15] J.P. Ma and B.H.J. McKellar, Preprint UM--P--96/55  
\par\noindent 
[16] G.P. Lepage et al., Phys. Rev. D46 (1992) 4052
\par\noindent
[17] P. Ko, J. Lee and H.S. Song, Phys. Rev. D54 (1996) 4312
\par\noindent
[18] M. Cacciari and M. Kr\"amer, Phys. Rev. Lett. 76 (1996) 4128 
\par\noindent 
[19] J. Amundson, S. Fleming and I. Maksymyk, Preprint 
MADTH-95-914, hep-ph/9601289 
\par\noindent
[20] P. Cho and A.K. Leibovich, Phys. Rev. D53 (1996) 6203 
\par\noindent 
[21] D. Allasia et al.; Phys. Lett. B258 (1991) 493 
\par\noindent
\ \ \ \ \ Ch. Mariotti, Nucl. Phys. A532 (1991) 437
\par\noindent
[22] A.H. Mueller, Phys. Lett. B308 (1993) 355
\par\noindent
\ \ \ \ \ X. Ji, Nucl. Phys. B448 (1995) 51 
\par\noindent
\vfil\eject
\centerline{\bf Figure Caption }
\par\vskip 20pt
\noindent
Fig.1A: The diagram appearing in the diagram expansion. This
diagram leads to contributions at leading twist and at higher 
twist. The wavy lines are for gluons. The lines connecting 
the black box with the upper part denote only 
momentum flows and contractions of color- and Lorentz-indices. 
\par\vskip5pt\noindent
Fig.1B: One of the two diagrams leads to contributions at next-to-leading
twist. In these diagrams the upper part is connected to  
the black box with three gluon lines. 
\par\vskip5pt\noindent
Fig.1C: The diagram leads to contributions at next-to-leading
twist. the momentum flows are given with $k_4=k_1+k_2-k_3$. 
\par\vskip5pt\noindent
Fig.2A: A diagram corresponds to the upper part in Fig.1A. Other
diagrams are obtained through permutations of gluon- and photon-lines. 
The line is for heavy quark. The black dot denotes the nonperturbative
transition into a quarkonium. The thick line is for quarkonium.  
\par\vskip5pt\noindent
Fig.2B: A diagram corresponds to the upper part in Fig.1B. Other
diagrams are obtained through permutations of gluon- and photon-lines.
\par\vskip 5pt\noindent
Fig.2C: A diagram corresponds to the upper part in Fig.1C. Other
diagrams are obtained through permutations of gluon- and photon-lines.
\par\vskip 5pt\noindent
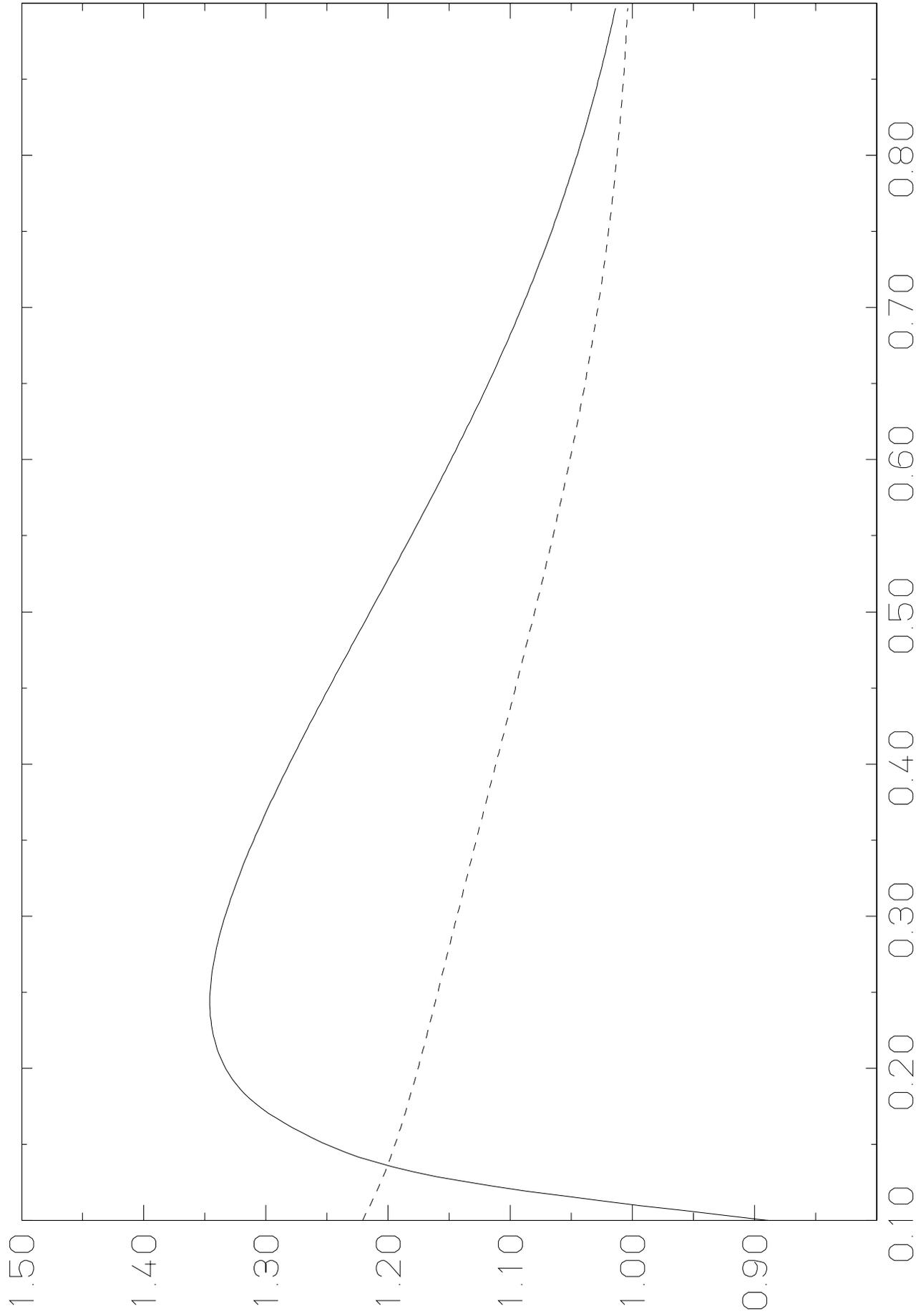
Fig.3: The quantity $\rho$ as function of $z$ defined in Eq.(3.20). 
The line is for $\hat s =(10)^2$GeV, the dashed one is for 
$\hat s=(30)^2$GeV. 
\par\vskip 5pt\noindent
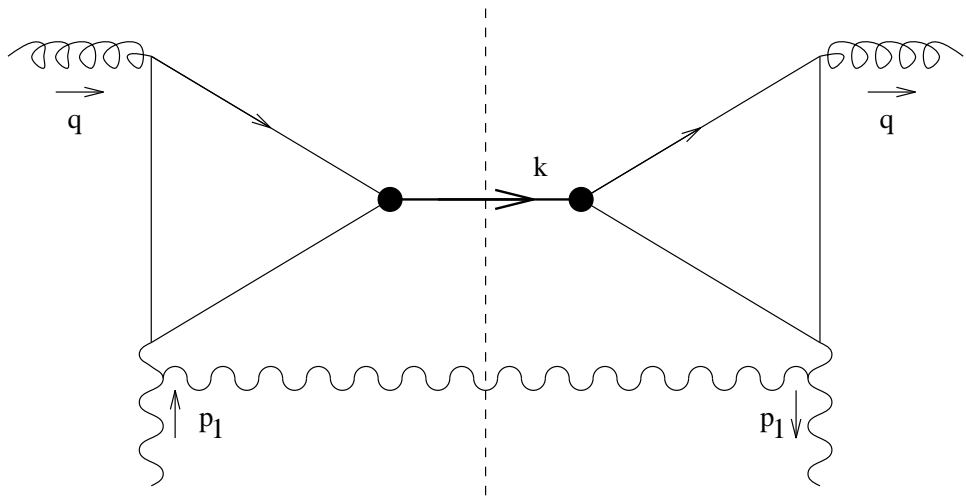
Fig.4: One of possible diagrams  in addition to those 
in Fig.2A contributes to 
the production rate through color-octet states 
\vfil\eject
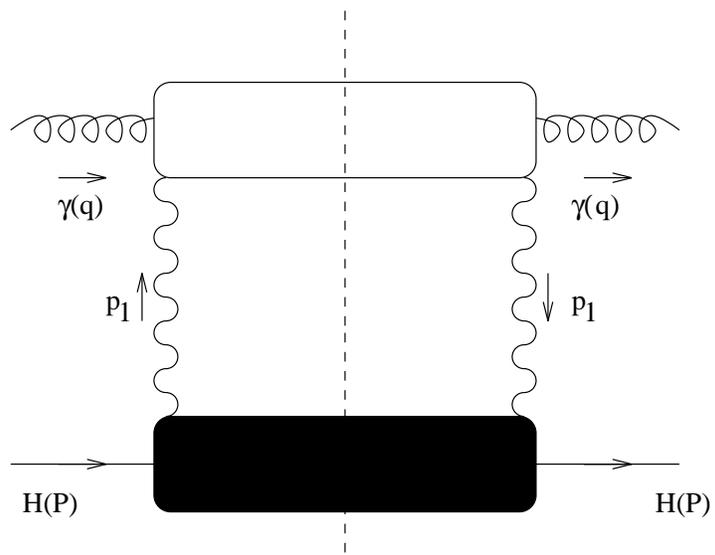
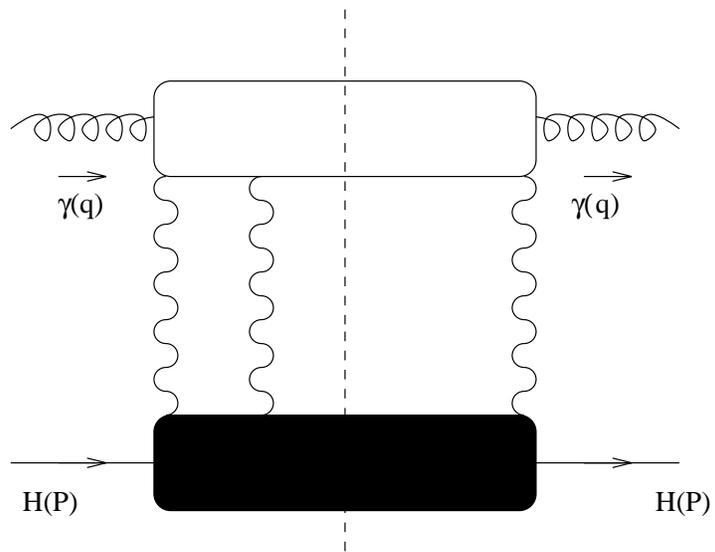
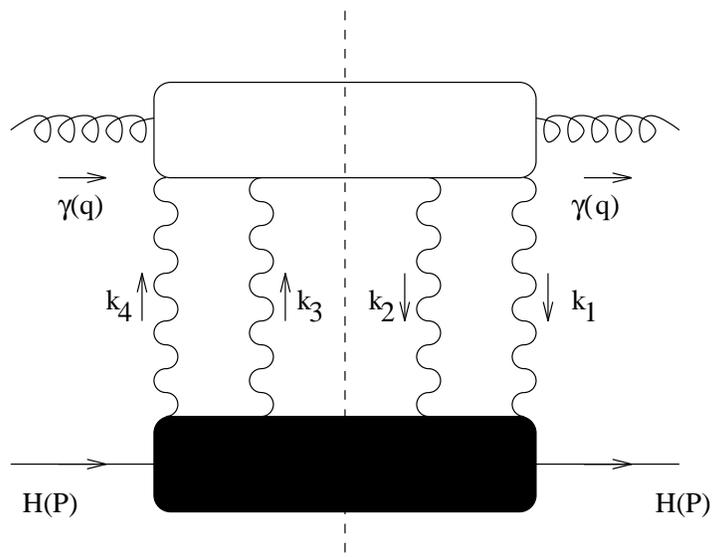
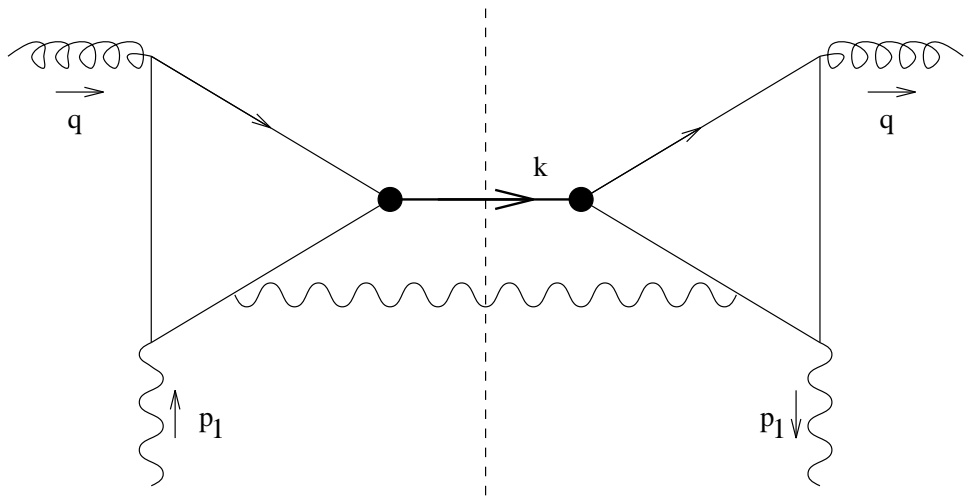
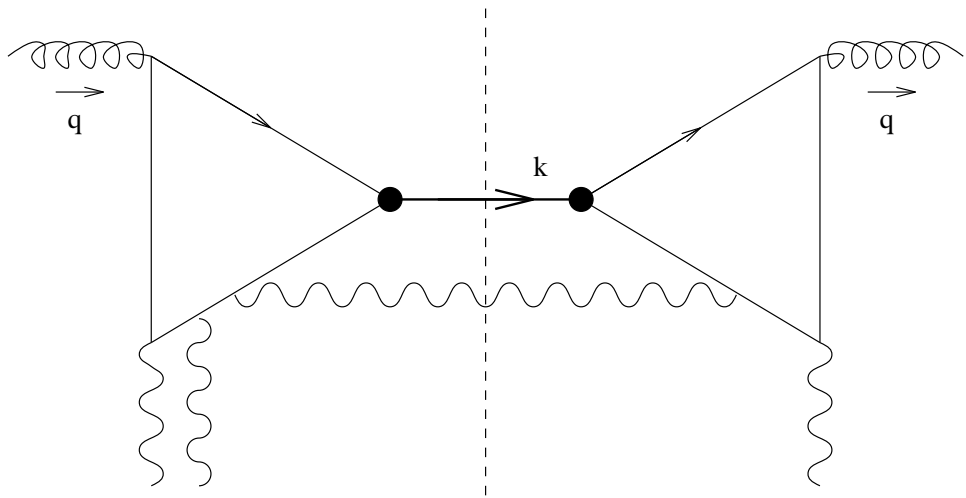
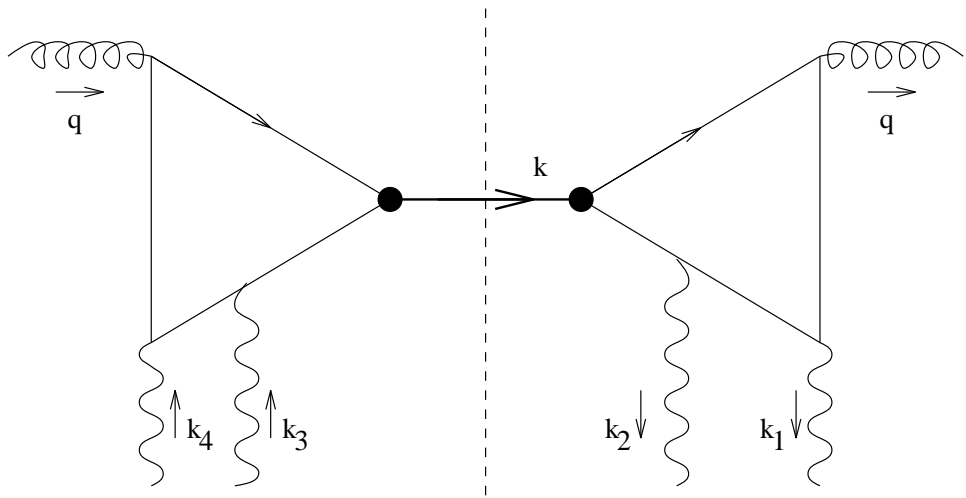
\end